\documentclass[aps,prl,twocolumn,showpacs,superscriptaddress,groupedaddress]{revtex4-1}
\usepackage{amsmath}
\usepackage{amsfonts}
\usepackage{graphicx}
\usepackage{caption}

\begin{document}

\title{Correlated random walks caused by dynamical wavefunction collapse}
\date{\today}
\author{D.~J.~Bedingham}
\email{daniel.bedingham@philosophy.ox.ac.uk}
\affiliation{Faculty of Philosophy, University of Oxford, OX2 6GG, United Kingdom.}
\author{H.~Ulbricht}
\email{h.ulbricht@soton.ac.uk}
\affiliation{School of Physics and Astronomy, University of Southampton, SO17 1BJ, United Kingdom.}

\begin{abstract}
Wavefunction collapse models modify Schr\"odinger's equation so that it describes the collapse of a superposition of macroscopically distinguishable states as a dynamical process. This provides a basis for the resolution of the quantum measurement problem. An additional generic consequence of the collapse mechanism is that it causes particles to exhibit a tiny random diffusive motion. Here it is shown that for the continuous spontaneous localization (CSL) model---one of the most well developed collapse models---the diffusions of two sufficiently nearby particles are positively correlated. An experimental test of this effect is proposed in which random displacements of pairs of free nanoparticles are measured after they have been simultaneously released from nearby traps. The experiment must be carried out at sufficiently low temperature and pressure in order for the collapse effects to dominate over the ambient environmental noise. It is argued that these constraints can be satisfied by current technologies for a large region of the viable parameter space of the CSL model. The effect disappears as the separation between particles exceeds the CSL length scale. The test therefore provides a means of bounding this length scale.
\end{abstract}
\maketitle

Dynamical wavefunction collapse models \cite{REP1,REP2} provide a unified description of quantum dynamics encompassing both unitary evolution and state reduction. Typically the standard Schr\"odinger dynamics are extended so that the state behaves stochastically and in such a way that certain superposition states are unstable. The most prominent collapse model is the continuous spontaneous localization (CSL) model \cite{CSL1,CSL2} in which a superposition of quasi-localized matter states will collapse at a rate which increases with the mass of the object. This results in the rapid collapse of macroscopically distinguishable superposition states whilst micro states are little affected.

Since the CSL model involves a modification of the Schr\"odinger equation, it makes predictions which are in conflict with standard quantum theory. It is therefore possible to experimentally test CSL. A direct way to do this is to try to observe quantum interference for objects of increasing mass. Detailed studies show that there should be a characteristic loss of fringe visibility as the mass increases beyond a sufficient size (see Refs \cite{ARNDT,MIRR,KIP}). Another observable effect of CSL is a random diffusive motion of particles. An isolated object will undergo a random walk induced by the collapse mechanism. As shown in Ref.\cite{CP} this can dominate over environmental effects at sufficiently low temperature and pressure.

Here we consider a situation with two non-interacting particles. We will examine how the particles behave as a result of the CSL dynamics and show that there is a potentially measurable effect whereby the diffusions undergone by each particle are correlated. A key feature of the CSL model is that the localization mechanism acts on the total smeared mass density state rather than individually on each particle. This suggests that for two nearby particles, the diffusive behaviour caused by the localization mechanism will be correlated. This should be apparent in the joint spatial probability distribution. In what follows we shall solve the two-particle CSL master equation to determine the behaviour precisely. 

\begin{figure}[h]
        \begin{center}
        	\includegraphics[width=5.0cm]{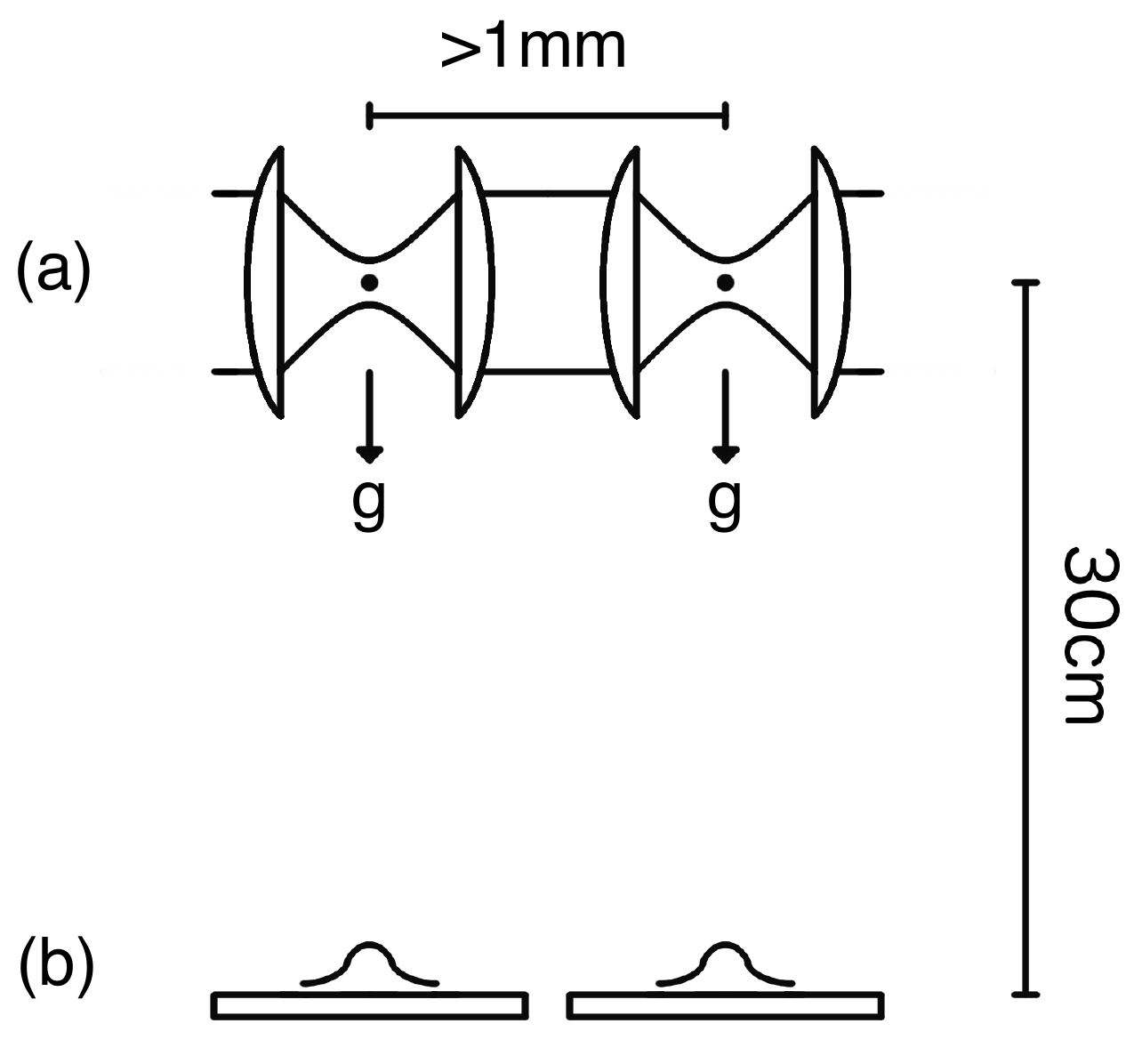}
        \end{center}
\caption{(a) Laser light creates adjacent harmonic traps for two nanoparticles. The particles are simultaneously released and fall $30$cm. The position of each particle is recorded with $10$nm accuracy at (b).}
\label{F0}
\end{figure}

Our proposed experiment is shown schematically in Fig.\ref{F0}. A pair of nanoparticles are trapped side by side using laser light. The particles are simultaneously released and allowed a brief period of free fall. The positions of each of the particles is then measured by light (Rayleigh) scattering.

We assume that the two particles are non-interacting and of equal mass. We also assume that the two particles are identical bosons although this is not crucial to our argument---the same conclusion holds if the particles are non-identical. We could also regard the particles as composite objects made up of many constituent particles satisfying the CSL dynamics. This case is more complicated but it can be shown \cite{CSL2} that the CSL dynamics can be applied to the composite object as a whole. Furthermore, any environmental particles involved in the CSL dynamics but not part of the system of interest can be traced away without effect.

There are two fixed parameters in the CSL model which are treated as fundamental constants. These are the localization rate $\lambda$ and the localization length scale $1/\sqrt{\alpha}$. We consider a regime in which the localization length scale is large compared to the length scale defining the spatial extent of the system. The original estimates of these parameters by Ghirardi, Rimini and Weber (GRW) \cite{GRW} are $\lambda = 10^{-16} {\rm s^{-1}}$ and $1/\sqrt{\alpha} = 10^{-7} {\rm m}$. However, these values are not definitive. A recent study of the valid regions of the $\lambda$-$\alpha$ parameter space consistent with experiment puts no upper bound limit on the CSL length scale \cite{TUM2}. 

The CSL master equation is given by
\begin{align}
\frac{\partial\hat{\rho}_t}{\partial t} = -\frac{i}{\hbar}\left[\hat{H},\hat{\rho}_t \right] 
- \frac{\lambda}{2}\int dx\left[\hat{M}(x),\left[ \hat{M}(x),\hat{\rho}_t \right]\right],
\label{Mop}
\end{align}
with density operator $\hat{\rho}_t$ and free Hamiltonian $\hat{H}$. The smeared mass density operator is given by
\begin{align}
\hat{M}({x}) = \left( \frac{\alpha}{\pi}\right)^{3/4} \frac{m}{m_0} \int d{y}\;e^{-\alpha({x}-{y})^2/2}\; \hat{a}^{\dagger}({y})\hat{a}({y}),
\label{M}
\end{align}
where $m_0$ is the nucleon mass (used to set the mass scale), and the field annihilation and creation operators $\hat{a}({x})$ and $\hat{a}^{\dagger}({x})$ satisfy
\begin{align}
[\hat{a}({x}),\hat{a}^{\dagger}({y})] =\delta({x}-{y}).
\end{align}
The two-particle state is represented by
\begin{align}
|\psi\rangle = \int d{x_1}dx_2\; \psi({x_1},x_2) \frac{1}{\sqrt{2}}\hat{a}^{\dagger}({x_1})\hat{a}^{\dagger}({x_2})|0\rangle,
\end{align}
where $1$ and $2$ label the two particles and $\psi$ is a symmetric wavefunction. In this representation improper position eigenstates are given by
\begin{align}
|x_1,x_2\rangle = \frac{1}{\sqrt{2}}\hat{a}^{\dagger}({x_1})\hat{a}^{\dagger}({x_2})|0\rangle,
\end{align}
and the two-particle density matrix is represented in coordinate space as
\begin{align}
\rho_t(x_1,y_1,x_2,y_2) = \langle x_1, x_2 | \hat{\rho}_t |y_1,y_2\rangle.
\end{align}
The coordinate space representation of Eq.(\ref{Mop}) for the two-particle state in the limit of large localization length is found to be 
\begin{align}
\frac{\partial}{\partial t}\rho_t &= \frac{i\hbar}{2m}\left(\frac{\partial^2}{\partial x_1^2} - \frac{\partial^2}{\partial y_1^2}\right)\rho_t + \frac{i\hbar}{2m}\left(\frac{\partial^2}{\partial x_2^2} - \frac{\partial^2}{\partial y_2^2}\right)\rho_t \nonumber\\
&-\frac{D}{\hbar^2}\left[(x_1-y_1)^2+(x_1-y_2)^2+(x_2-y_1)^2\right.\nonumber\\
&\quad\left.+(x_2-y_2)^2-(x_1-x_2)^2-(y_1-y_2)^2\right]\rho_t,
\label{M2}
\end{align}
where 
\begin{align}
D = \frac{\hbar^2 \lambda\alpha}{4} \left(\frac{m}{m_0}\right)^2.
\end{align}
Here, the limit of large localization length specifically means that $1/\sqrt{\alpha} \gg |x_i-y_i|$ for $i = 1,2$, and $1/\sqrt{\alpha} \gg |x_1-x_2|$. From here on we treat Eq.(\ref{M2}) in one dimension having traced out the other two. 

We now present a solution of Eq.(\ref{M2}) (further details of the techniques used can be found in Ref.\cite{DB}). The solution is represented in terms of a density matrix propagator $J$ (see Refs \cite{CALD,ANAS}) as
\begin{align}
&\rho_t(x_1,y_1,x_2,y_2)\nonumber\\
&=\int dx'_1dy'_1dx'_2dy'_2  J(x_1,y_1,x_2,y_2,t|x_1',y_1',x_2',y_2',0)\nonumber\\
&\quad\quad\quad\quad\quad\quad\quad\quad\times\rho_0(x'_1,y'_1,x'_2,y'_2).
\label{prop1}
\end{align}
We find that for Eq.(\ref{M2}), $J$ is given by
\begin{align}
J(x_1&,y_1,x_2,y_2,t|x_1',y_1',x_2',y_2',0)\nonumber\\
=&\left(\frac{m}{2\pi \hbar t}\right)^2 e^{ (im/2\hbar t) \sum_i\left[(x_i-x_i')^2-(y_i-y_i')^2 \right]}\nonumber\\
& \times e^{ -(Dt/3\hbar^2)\sum_{i,j}\left[ (x_i-y_j)^2+(x_i-y_j)(x_i'-y_j')+(x_i'-y_j')^2 \right] }\nonumber\\
& \times e^{ (Dt/3\hbar^2)\left[ (x_1-x_2)^2+(x_1-x_2)(x_1'-x_2')+(x_1'-x_2')^2 \right] }\nonumber\\
& \times e^{ (Dt/3\hbar^2)\left[ (y_1-y_2)^2\; +(y_1-y_2)(y_1'-y_2')\;+(y_1'-y_2')^2 \right] },
\label{prop2}
\end{align}
where $i,j = 1,2$. 

We consider an initial wave function of the form
\begin{align}
\psi(x_1,x_2) =& \frac{1}{\sqrt{4\pi\sigma^2}}e^{-(x_1-\mu)^2/4\sigma^2}e^{-(x_2+\mu)^2/4\sigma^2}
\nonumber \\
&+\frac{1}{\sqrt{4\pi\sigma^2}}e^{-(x_1+\mu)^2/4\sigma^2}e^{-(x_2-\mu)^2/4\sigma^2}.
\end{align}
This represents the particles residing in adjacent harmonic traps of width $\sigma$ and spaced by a distance $2\mu\gg\sigma$. The initial density matrix is
\begin{align}
\rho_0(x_1,y_1,x_2,y_2) = \psi(x_1,x_2) \psi^*(y_1,y_2).
\end{align}
We suppose that the particles are simultaneously released from the traps and allowed to diffuse freely. However, they remain sufficiently separated that there is negligible conventional interaction between them. Using Eqs (\ref{prop1}) and (\ref{prop2}) we have calculated the diagonal part of the density matrix 
\begin{align}
P_t(x_1,x_2) =\rho_t(x_1,x_1,x_2,x_2),
\end{align}
following a time period $t$ after their release. This represents the joint probability distribution for the subsequently measured positions of the two particles. The result is simplest when expressed in terms of the variables
\begin{align}
X = \frac{x_1+x_2}{2} \quad \text{and} \quad \xi = x_1-x_2,
\end{align}
where we find
\begin{align}
P_t(X,\xi) =&\frac{1}{8\pi\sigma_X\sigma_{\xi/2}} e^{-X^2/2\sigma^2_X}e^{-(\xi/2-\mu)^2/2\sigma^2_{\xi/2}} \nonumber \\
&+\frac{1}{8\pi\sigma_X\sigma_{\xi/2}} e^{-X^2/2\sigma^2_X}e^{-(\xi/2+\mu)^2/2\sigma^2_{\xi/2}} ,
\label{jointP}
\end{align}
with
\begin{align}
\sigma^2_X =& \frac{1}{2}\sigma^2 \left(\frac{4Dt^3}{3m^2\sigma^2}+\frac{\hbar^2t^2}{4m^2\sigma^4} +1\right),\label{s1} \\
\sigma^2_{\xi/2} =& \frac{1}{2}\sigma^2 \left(\frac{\hbar^2t^2}{4m^2\sigma^4} +1\right).
\label{s2}
\end{align}
The distribution consists of two peaks, one about $X=0,\xi/2=\mu$ and another about $X=0,\xi/2=-\mu$. Note that we have ignored a possible interference term between the two peaks in Eq.(\ref{jointP}) since we assume that they do not disperse enough to overlap. The feature that we are interested in is the shape of these peaks and in particular their rate of dispersion in the two directions $X$ and $\xi/2$. The $X$ and $\xi$ spreads of each of the peaks are defined by $\sigma_X$ and $\sigma_{\xi/2}$ respectively. We see that  the CSL parameter $D$ contributes to $\sigma_X$ whilst $\sigma_{\xi/2}$ is unaffected. Note that by setting $D=0$ (standard quantum mechanics) in Eqs (\ref{s1}) and (\ref{s2}) we find $\sigma^2_X =\sigma^2_{\xi/2}$. This reflects the fact that the two particles are behaving independently and their distributions are uncorrelated.

The physical reason for the difference between $\sigma_X$ and $\sigma_{\xi/2}$ when $D\neq 0$ is that the diffusive shifts in position of the two particles caused by the collapse mechanism will be positively correlated if the particles are closer together than the localization length scale. The system as a whole diffuses but $\xi = x_2-x_1$ is only affected by standard quantum dispersion.

Our result is clearly dependent on the fact that the localization length scale $1/\sqrt{\alpha} > 2\mu$, but in this limit the spreads in $X$ and $\xi/2$ do not depend on the separation $2\mu$. For $1/\sqrt{\alpha} < \mu$ the particles would behave independently under CSL. A null observation would therefore put a constraint on $\alpha$.

\begin{figure}[h]
        \begin{center}
        	\includegraphics[width=9.0cm]{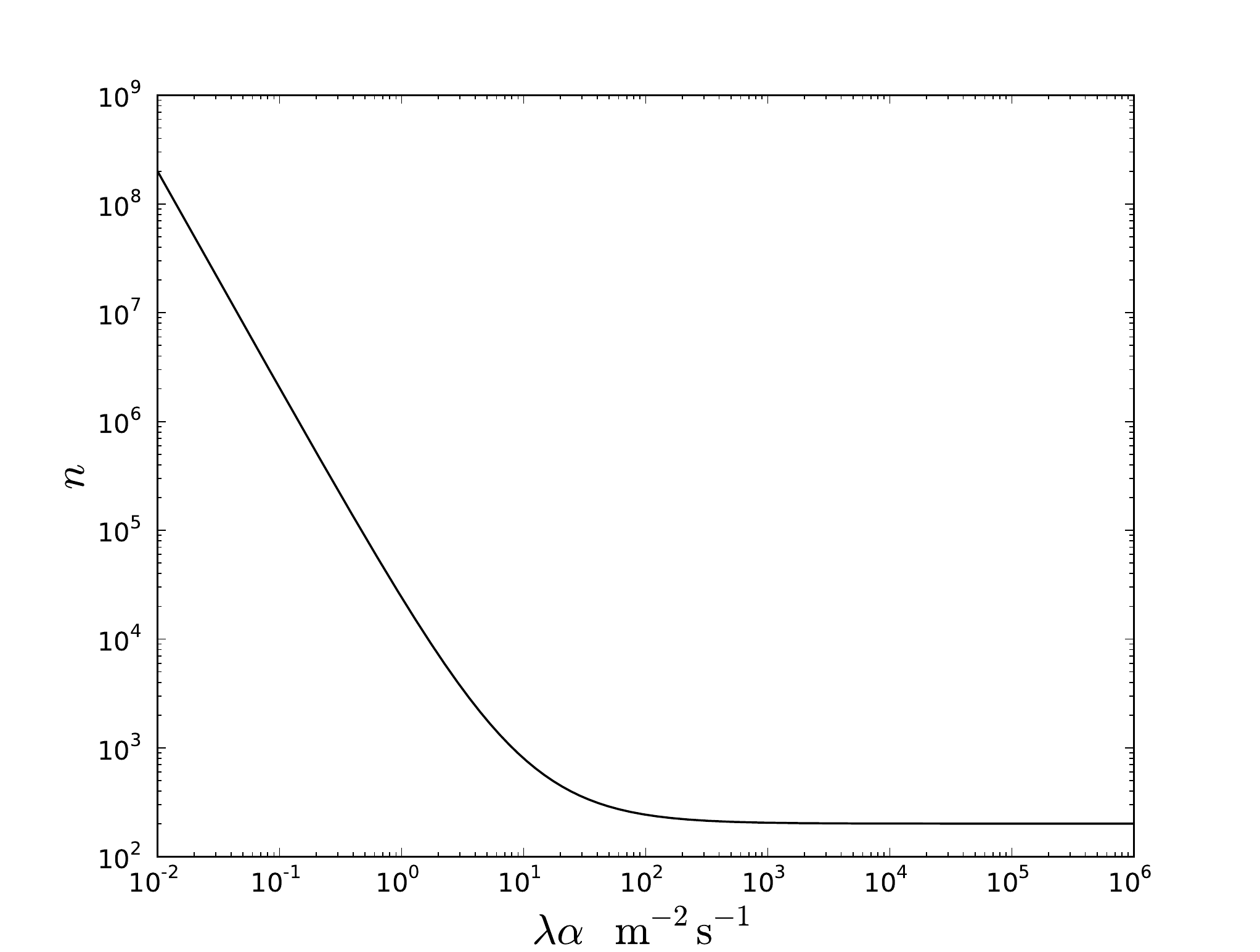}
        \end{center}
\caption{Minimum number of observations $n$ required to estimate variances $\sigma^2_X$ and $\sigma^2_{\xi/2}$ with sufficient accuracy to determine their expected difference $\sigma^2_{CSL}$.}
\label{F1}
\end{figure}

To observe the effect we must distinguish $\sigma^2_X$ and $\sigma^2_{\xi/2}$. We can estimate these variances by repeatedly measuring the displacements of two particles simultaneously dropped from nearby traps (see Fig.\ref{F0}). We assume that the traps are such that $\sigma = 10{\rm nm}$ and the particles are released for $t =0.25 {\rm s}$ corresponding to a drop of $30{\rm cm}$. We further assume that position measurements have an error of $\sigma_{err} = 10{\rm nm}$ which is normally distributed and independent of $x_i$. If $s^2_X$ is the unbiased estimate of $\sigma^2_X$ based on $n$ sets of position measurements, then the variance in $s^2_X$ is given by
\begin{align}
{\rm Var}[s^2_X] = \frac{2\left(\sigma_X^2 + \frac{1}{2}\sigma^2_{err}\right)^2}{n-1}.
\end{align} 
To be able to clearly observe the effect of CSL we demand that $\sqrt{{\rm Var}[s^2_X]}$ should be at least 10 times smaller than the difference
\begin{align}
\sigma^2_{CSL} = \sigma^2_X -\sigma^2_{\xi/2} = \frac{2Dt^3}{3m^2}.
\end{align}
Taking the mass of the particles to be $10^9 {\rm amu}$ such that $\hbar^2t^2/4m^2\sigma^4\ll1$, this results in the following constraint on $n$:
\begin{align}
n>2\left(\frac{100}{\lambda\alpha} + 10\right)^2 +1.
\end{align}
This is shown in Fig.\ref{F1}.

Now we consider the effects of an environment. We suppose that the particles are confined within a vacuum chamber and consider the constraints placed on temperature and pressure by the condition that the collapse effects should be dominant. We consider environmental noise contributions both from radiation and molecular collisions. 

\begin{figure}[h]
        \begin{center}
        	\includegraphics[width=9.0cm]{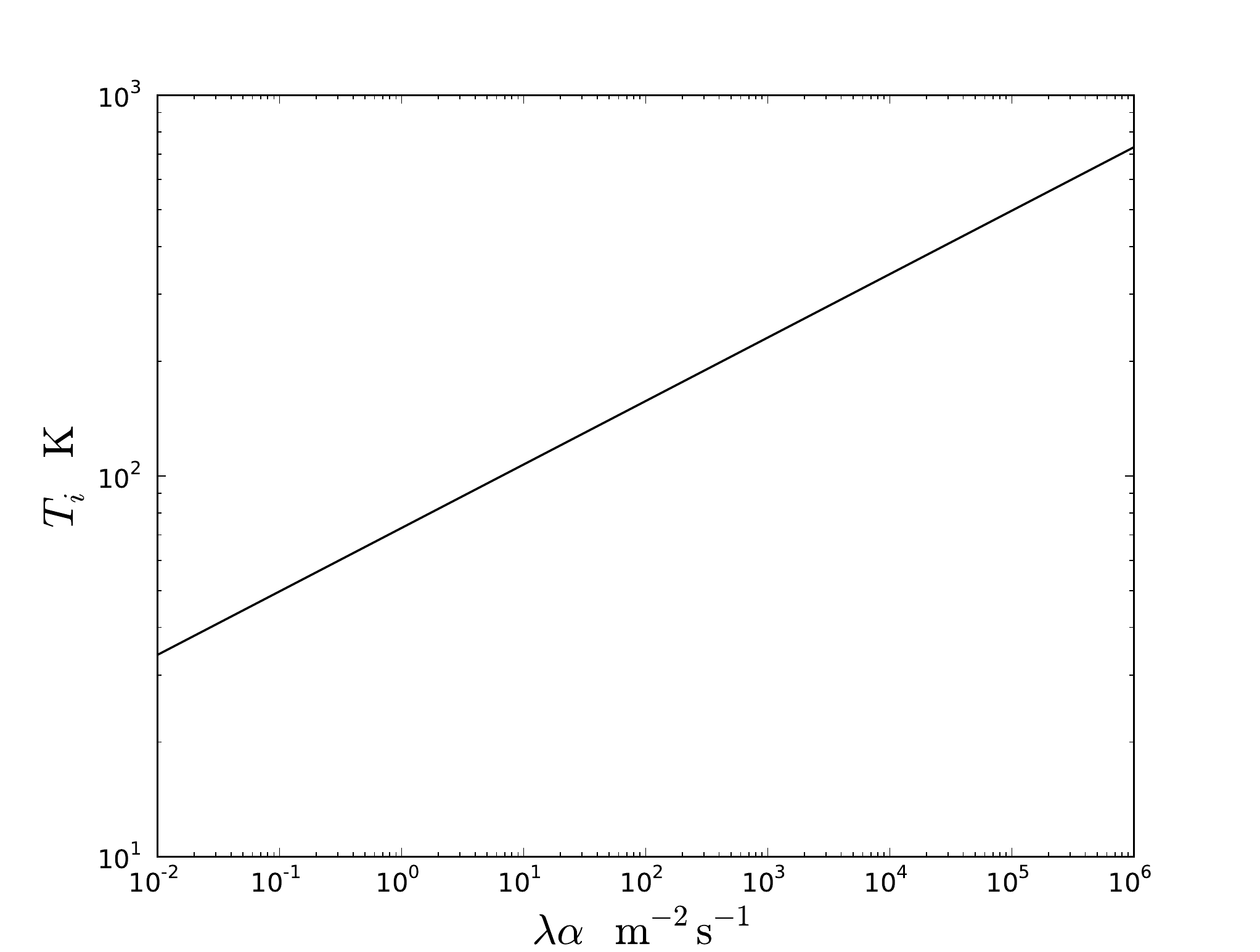}
        \end{center}
\caption{Maximum internal temperature of particles required to prevent thermal diffusion by radiation.}
\label{F2}
\end{figure}

It is shown in Ref.\cite{BERA} that the dominant contribution to the thermal diffusion by radiation is due to recoil from emission of photons. There it is also shown that the variance in displacement of a bulk sphere of radius $R$ and density ${\cal D}$ due to the emission of radiation is given (in SI units) by
\begin{align}
\sigma^2_{RAD}  = 4.0\times 10^{-43} {\cal D}^{-2} R^{-3} T_i^6 t^3,
\end{align}
where $T_i$ is the internal temperature of the bulk object. By demanding that $\sigma^2_{CSL} > 10\times \sigma^2_{RAD}$ we can constrain the internal temperature of the particles. Using ${\cal D} = 10^3 {\rm kgm^{-3}}$ and $R = 10^{-7}{\rm m}$, we find
\begin{align}
T_i < 73(\lambda\alpha)^{1/6}.
\end{align}
The maximum internal temperature is shown in Fig.\ref{F2}.

\begin{figure}[h]
        \begin{center}
        	\includegraphics[width=9.0cm]{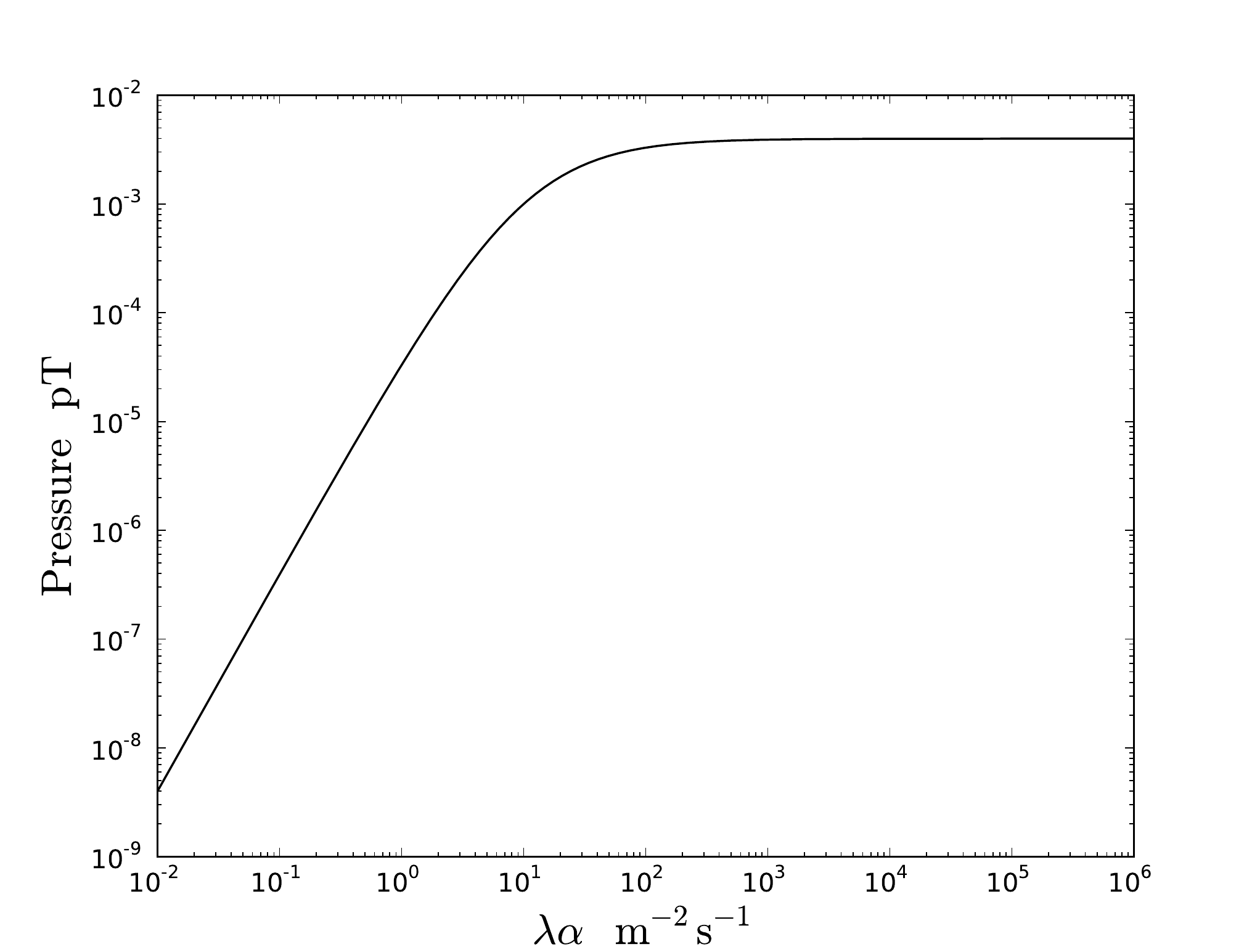}
        \end{center}
\caption{Maximum ambient pressure required to prevent diffusion due to molecular collisions.}
\label{F3}
\end{figure}

The mean time between molecule-sphere collisions in the impact regime is given in seconds by \cite{CP}
\begin{align}
\tau_c\simeq 2 \left(\frac{T_e}{T_0}\right)^{1/2} P^{-1},
\end{align}
where $P$ is the pressure given in picoTorr, $T_e$ is the external temperature, and $T_0=300{\rm K}$ is room temperature. We make the conservative assumption that $\tau_c$ should be $10$ times greater than the total experiment time $n\times t$. Taking $T_e/T_0 \sim 1$ this results in 
\begin{align}
P < 0.8\left[2\left(\frac{100}{\lambda\alpha} + 10\right)^2 +1\right]^{-1}.
\end{align}
The maximum ambient pressure in the vacuum chamber is shown in Fig.\ref{F3}.

\begin{figure}[h]
        \begin{center}
        	\includegraphics[width=9.0cm]{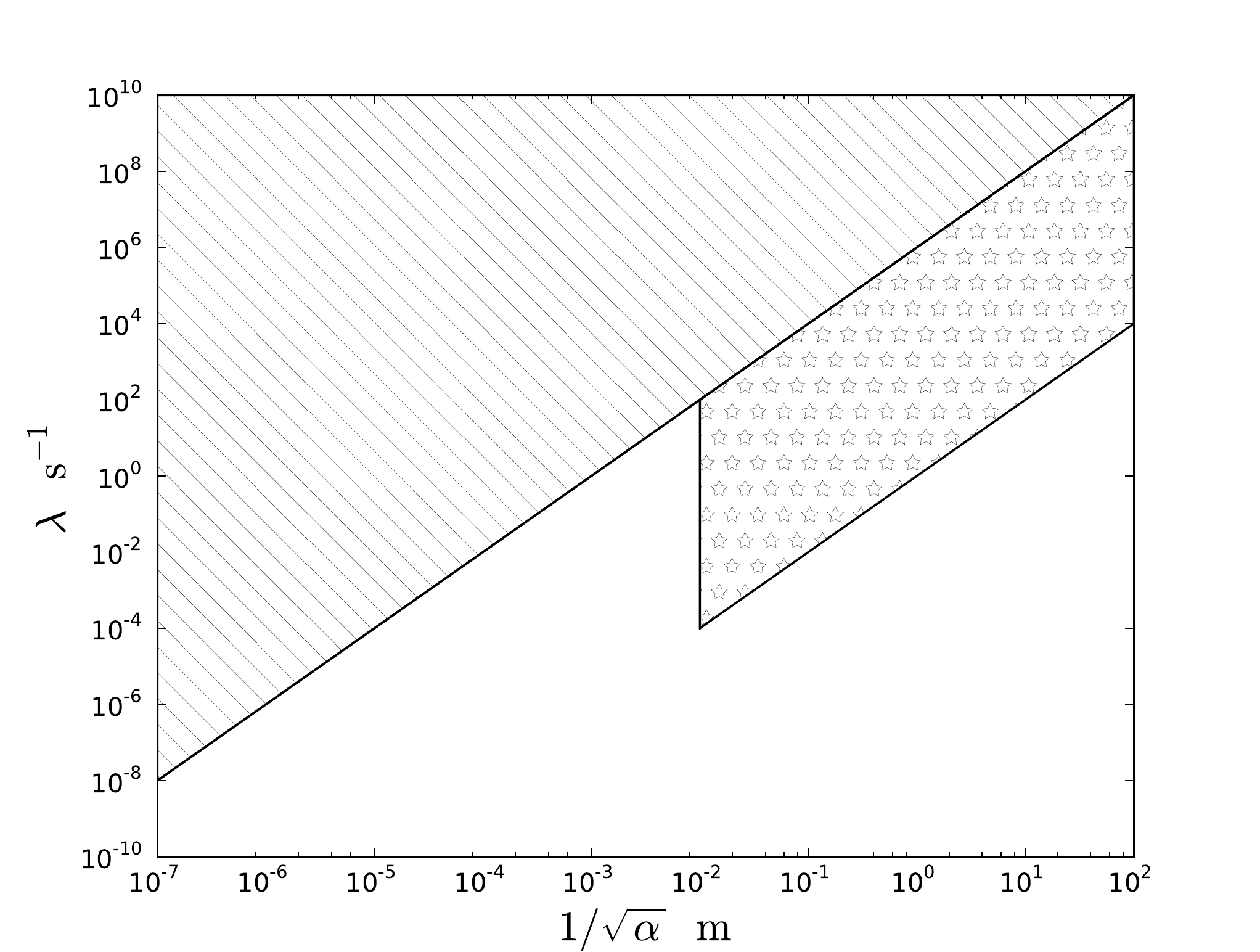}
        \end{center}
\caption{CSL parameter diagram. Hatched area shows the region of $\lambda$-$\alpha$ parameter space currently ruled out by diffraction experiments. Starred area shows the region of parameter space in which the correlated random walks are experimentally accessible.}
\label{F4}
\end{figure}

We conclude that an experiment can be performed to test for $\lambda\alpha$ as low as $1{\rm m^{-2}s^{-1}}$. This requires $n = 24000$; $T_e = 73 {\rm K}$ and $P = 3.3\times10^{-17} {\rm Torr}$. We assume a separation between traps of at least $1$mm (to avoid dispersion forces/gravity effects) so that the correlation effect should be present for  $1/\sqrt{\alpha} \geq 1{\rm cm}$. The accessible region of $\lambda$-$\alpha$ parameter space is shown in Fig.\ref{F4}. Also shown is the region of parameter space currently ruled out by diffraction experiments involving particles of mass $10^{5}{\rm amu}$ \cite{DIFR}.

In summary, we have demonstrated an effect of the CSL model in which the diffusive behaviour of two sufficiently nearby particles is correlated. If one particle is found to have randomly moved in one direction as a result of the collapse mechanism, the other particle is more likely to have moved in the same direction. We propose attempting to observe this effect as a test of CSL against standard quantum theory and as a specific test of the CSL length scale. The experiment is possible with today's technology.

DJB was supported by EPSRC Research Grant No.~EP/J008060/1 and a grant from the Templeton World Charity Foundation. HU acknowledges support by the EPSRC (EP/J014664/1), the Foundational Questions Institute (FQXi) through a Large Grant, and by the John F Templeton foundation (grant 39530). We would like to thank Jonathan Halliwell for helpful discussions and comments.

\end{document}